\documentclass[journal]{IEEEtran}
\usepackage{textcomp}
\usepackage{xcolor}
\usepackage{graphicx}
\usepackage{amsmath}
\usepackage{amssymb}
\usepackage{amsfonts}
\usepackage{epsfig}
\usepackage{epstopdf}
\usepackage{hhline}
\usepackage{stfloats}
\usepackage{cite}
\usepackage{algorithm}
\usepackage{algorithmic}
\usepackage[CJKbookmarks=true,
bookmarksnumbered=true,
bookmarksopen=true]{hyperref}
\usepackage{multirow}
\usepackage{subfigure}

\begin{document}
\title{Communication-efficient Personalized Federated Edge Learning for Massive MIMO CSI Feedback}
\author{Yiming~Cui, \emph{Graduate Student Member}, \emph{IEEE},
Jiajia~Guo, \emph{Member}, \emph{IEEE},
Chao-Kai~Wen, \emph{Fellow}, \emph{IEEE}, and~Shi~Jin, \emph{Fellow}, \emph{IEEE}

\thanks{

Y. Cui, J. Guo, and S. Jin are with the National Mobile Communications Research Laboratory, Southeast University, Nanjing 210096, China (e-mail: cuiyiming@seu.edu.cn; jiajiaguo@seu.edu.cn; jinshi@seu.edu.cn).
	
C.-K. Wen is with the Institute of Communications Engineering, National Sun Yat-sen University, Kaohsiung 80424, Taiwan (e-mail: chaokai.wen@mail.nsysu.edu.tw).}

}		

\maketitle

\begin{abstract} 
Deep learning (DL)-based channel state information (CSI) feedback has garnered significant research attention in recent years. However, previous research has overlooked the potential privacy disclosure problem caused by transmitting CSI datasets during the training process. In this study, we introduce a federated edge learning (FEEL)-based training framework for DL-based CSI feedback. This approach differs from the conventional centralized learning (CL)-based framework, where the CSI datasets are collected at the base station (BS) before training. Instead, each user equipment (UE) trains a local autoencoder network and exchanges model parameters with the BS. This approach provides better protection for data privacy compared to CL. To further reduce communication overhead in FEEL, we quantize the uplink and downlink model transmission into different bits based on their influence on feedback performance. Additionally, since the heterogeneity of CSI datasets among different UEs can degrade the performance of the FEEL-based framework, we introduce a personalization strategy to enhance feedback performance. This strategy allows for local fine-tuning to adapt the global model to the channel characteristics of each UE. Simulation results indicate that the proposed personalized FEEL-based training framework can significantly improve the performance of DL-based CSI feedback while reducing communication overhead.

\end{abstract}

\section{Introduction}

Massive multiple-input multiple-output (MIMO) technology is a fundamental component of the fifth generation (5G) and is anticipated to continue playing a pivotal role in forthcoming generations of cellular communication systems. A typical massive MIMO system consists of a base station (BS) equipped with a substantial number of antennas. To fully exploit the benefits of massive MIMO, such as spatial diversity and multiplexing gains, acquiring accurate downlink channel state information (CSI) is essential \cite{lu2014overview}. In time division duplex mode, the BS can obtain downlink CSI from uplink CSI due to bi-directional reciprocity. However, reciprocity doesn't hold in frequency division duplex (FDD) mode. In FDD, user equipment (UE) needs to transmit downlink CSI back to the BS through feedback links, leading to increased communication overhead, especially with a large number of antennas at the BS \cite{liang2016fdd}.

Various conventional methods have been proposed to mitigate feedback overhead, mainly including codebook-based and compressive sensing (CS)-based approaches. The codebook-based method has been standardized in current communication systems, such as the Type I/II codebook in 5G New Radio \cite{TYPE12}. This approach involves the UE selecting codewords from the shared codebook with the BS and transmitting the corresponding index to the BS. The BS then recovers the channel using the codebook \cite{love2008overview}. Nevertheless, codebook-based methods are only suitable for regular antenna arrays and the feedback overhead scales approximately with the antenna count. Moreover, codebook design becomes intricate as the number of antennas grows. In contrast, the CS-based feedback method compresses CSI using random projection into low-dimensional codewords, which are reconstructed at the BS using iterative algorithms like BM3D-AMP \cite{metzler2016denoising}. However, the efficacy of the CS-based method relies heavily on the sparsity of channel matrices, which may not always hold true in real communication scenarios. Hence, more efficient CSI feedback methods are needed to alleviate feedback overhead.

To address the limitations of the aforementioned methods and enhance feedback performance, deep learning (DL) was introduced to CSI feedback in \cite{wen2018deep}. This DL-based method employs an autoencoder neural network to autonomously learn how to compress and reconstruct CSI from a large dataset of CSI samples. In comparison to codebook-based and CS-based methods, the DL-based approach significantly boosts feedback performance by leveraging environmental knowledge extracted from CSI samples to aid compression and reconstruction. Furthermore, the DL-based CSI feedback algorithm is highly parallel and can be accelerated using specialized computational devices like graphics processing units and tensor processing units. Additionally, DL-based CSI feedback can accommodate various antenna array configurations without depending on channel sparsity. These advantages have garnered significant attention from academia and industry for DL-based CSI feedback. In 2021, CSI feedback enhancement was selected as a representative use case in the 3rd Generation Partnership Project's (3GPP) study item on artificial intelligence and machine learning for the new radio air interface \cite{213599} in Release 18. The DL-based CSI feedback approach in \cite{wen2018deep} is considered a typical method in this context and is widely recognized as a promising solution for future communication systems. Numerous research efforts have expanded upon DL-based CSI feedback \cite{guo2022overview}, which are briefly summarized in Table \ref{tab:relatedwork}.

\begin{table}[t]
    \renewcommand{\arraystretch}{1.0}
    \centering
    \caption{Recent research on DL-based CSI feedback}
    \begin{tabular}{|l|l|} 
    \hline
         \textbf{Category} & \textbf{Related research} \\
         \hline
         \multirow{3}{*}{Novel network architectures} & attention mechanism \cite{cai2019attention,zhang2022attention}\\
         & multiple-resolution designs \cite{lu2020multi}\\
         & non-local blocks \cite{yu2020ds}\\
         & generative networks \cite{tolba2020massive}\\
         \hline
         \multirow{4}{*}{Different types of correlation} & temporal correlation \cite{wang2018deep}\\
         & frequency correlation \cite{wang2021compressive}\\
         & bi-directional correlation \cite{liu2019exploiting}\\
         & correlation between nearby UEs \cite{mashhadi2020distributed} \\
         \hline
         \multirow{3}{*}{Joint design with other modules} & channel estimation \cite{chen2020deep}\\
         & UE selection \cite{mao2021dl} \\
         & beamforming \cite{guo2020deep} \\
         \hline
         \multirow{4}{*}{Deployment problems} & bit-stream generation \cite{guo2020convolutional} \\
         & implicit feedback \cite{chen2021deep} \\
         & transmission errors \cite{ye2020deep} \\
         & complexity reduction \cite{lu2022binarized} \\
    \hline

    \end{tabular}
    \label{tab:relatedwork}
\end{table}

Firstly, various novel neural network architectures, originally used in computer vision, have been adapted for autoencoder design, capitalizing on the resemblance between CSI matrices and images. These architectures encompass attention mechanisms \cite{cai2019attention,zhang2022attention}, multi-resolution designs \cite{lu2020multi}, non-local blocks \cite{yu2020ds}, generative networks \cite{tolba2020massive}, and more.

Secondly, different forms of correlation have been incorporated into CSI feedback to further enhance its performance. For instance, temporal correlation between CSI within a coherence time is leveraged using long short-term memory modules \cite{wang2018deep}. The correlation between the magnitude of uplink and downlink channels is exploited in \cite{liu2019exploiting}. The correlation between channels in adjacent subcarriers is harnessed through frequency domain downsampling in \cite{wang2021compressive}. Additionally, the spatial correlation among neighboring UEs is introduced in \cite{mashhadi2020distributed}. Moreover, CSI feedback has been jointly designed with other modules, including channel estimation \cite{chen2020deep}, beamforming \cite{guo2020deep}, UE selection \cite{mao2021dl}, and more.

Last but not least, practical deployment of DL-based CSI feedback is addressed in several research works. Deployment topics include, but are not limited to, bit-stream generation \cite{guo2020convolutional}, implicit feedback \cite{chen2021deep}, handling transmission errors \cite{ye2020deep}, and reducing complexity \cite{lu2022binarized}. Prior research on DL-based CSI feedback has mostly relied on offline training and assumed that access to the CSI datasets in the target environment is readily available. However, this assumption is rarely met in real-world communication systems. An intuitive approach to address this challenge is to train autoencoder networks using datasets collected from other environments or generated through channel simulation software. However, such datasets often fail to accurately represent the true channel distribution, leading to suboptimal feedback performance. Therefore, a practical online training method is needed to implement DL-based CSI feedback in real systems.

Traditionally, two approaches, individual learning (IL) and centralized learning (CL), have been considered for online training in DL-based CSI feedback. Under the IL approach, each UE independently trains a private autoencoder network using its local CSI dataset for CSI compression and reconstruction. However, the observations of the local datasets are often limited, leading to suboptimal generalization capabilities. Moreover, storage limitations may prevent UEs from gathering sufficient CSI samples to train an effective autoencoder network, resulting in subpar feedback performance. On the other hand, in the CL approach, a global autoencoder network is trained for all UEs. UEs transmit their local CSI datasets to a central server, which aggregates these datasets to train the global autoencoder. The generalization capability of the global autoencoder is usually superior to that of private autoencoder networks trained through IL due to the larger dataset. However, CL involves transmitting CSI datasets, which raises privacy concerns as the location and habits of UEs can be inferred from the data \cite{ma2019wifi,wang2016csi,wu2012csi}. This raises the possibility of privacy breaches.\footnote{Although originally intended for BS-directed CSI feedback in FDD systems, DL-based algorithms might involve other computation nodes to complete neural network training, potentially posing privacy risks.}

To simultaneously achieve strong generalization and safeguard privacy, online training in DL-based CSI feedback can adopt a distributed training method known as federated learning (FL) \cite{mcmahan2017communication}. Unlike CL, where all datasets are centralized at a server, FL maintains datasets locally to ensure data privacy. For model training, the central server schedules UEs to conduct local training and exchanges model parameters or gradient information with them. In discussions about AI-based CSI feedback enhancement in 3GPP, FL is regarded as a promising solution for autoencoder network training by many companies \cite{2205556}. Notably, unlike common applications like next-word prediction \cite{hard2018federated}, healthcare\cite{antunes2022federated}, and recommendation systems \cite{yang2020federated}, where the central server is usually in the cloud, for physical layer applications, the central server is typically an edge server located near the BS. This is because UEs in FL are within the same cell \cite{zhu2020toward}. Consequently, to emphasize the interaction between the BS and UEs, the term ``federated edge learning'' (FEEL) is often used to refer to FL in physical layer applications \cite{ren2020scheduling,zhu2019broadband,tak2020federated}. In comparison to conventional cloud server-based FL that serves a broader range of UEs, FEEL prioritizes physical layer applications with lower transmission latency and stronger dataset correlation among UEs in the same cell. The FEEL-based training framework has been applied to various physical layer tasks, including DL-based channel estimation \cite{elbir2021federated}, detection \cite{yang2021federated}, beamforming \cite{elbir2020federated}, beam management \cite{xue2023beam}, channel prediction \cite{hou2021federated}, and more.

This study introduces a communication-efficient FEEL-based training framework for DL-based CSI feedback. In this framework, UEs collaborate to train a shared global autoencoder for CSI feedback under the coordination of the BS. Each UE first trains its autoencoder with local CSI datasets and then uploads gradient information to the BS. The BS aggregates the gradients to update the global autoencoder. Notably, the BS and UEs exchange only model parameters (or gradient information), preserving the privacy of UEs by not sharing CSI datasets. To reduce the communication overhead stemming from model parameter transmission, a parameter quantization method is introduced to FEEL. This method quantizes gradients during uplink transmission and the global model during downlink transmission into varying numbers of bits based on their impact on feedback performance. However, this FEEL-based training framework faces challenges in dealing with significant data heterogeneity in DL-based CSI feedback. While UEs in the same cell share some common channel features to a degree, they typically operate in diverse environments with varying channel distributions. Thus, to enhance performance in heterogeneous settings, an alternative approach is proposed to allow each UE to adapt the global model using its local CSI dataset to suit its surrounding environment. This approach, commonly referred to as personalization \cite{tan2022towards}, is incorporated into the FEEL-based training framework. Finally, the communication-efficient FEEL-based training framework and the personalization strategy are evaluated using CSI datasets generated by the QuaDRiGa channel generation software \cite{jaeckel2014quadriga}.

This work contributes in the following ways:
\begin{itemize}
    \item \textbf{Privacy-preserving}: A FEEL-based training framework is proposed for DL-based CSI feedback, effectively addressing privacy concerns while achieving performance comparable to the CL-based approach.

    \item \textbf{Communication-efficient}: To further minimize communication overhead in the FEEL-based framework, a parameter quantization strategy is introduced, focusing on reducing the bottleneck of model transmission in FEEL.

    \item \textbf{High-performance}: Unlike previous FEEL-based physical layer designs that utilize only the global model, this study introduces a fine-tuning-based personalization strategy, substantially improving feedback performance within the FEEL-based framework.
\end{itemize}

The notations used in this paper are provided below: Bold letters denote vectors and matrices, while non-bold letters indicate scalars. The notation $(\cdot)^H$ denotes the Hermitian transpose operation. $\mathbb{C}^{s\times t}$ ($\mathbb{R}^{s\times t}$) represents a complex (real) space with dimensions $s$ by $t$. $\| \cdot \|_2$ signifies the Euclidean norm. $|\cdot|$ represents the cardinality of a set.

\section{System Model}
This section introduces the system model. We begin by presenting the channel model and signal model, followed by a discussion of the DL-based CSI feedback framework.

\subsection{Channel Model and Signal Model}
We consider a single-cell FDD system employing massive MIMO. The BS is equipped with a uniform linear array comprising $N_{\rm t} \gg 1$ antennas, while the UE has a single antenna. The system utilizes orthogonal frequency division multiplexing (OFDM) across $N_{\rm c}$ subcarriers. The channel vector on the $n$-th subcarrier $\widetilde{\mathbf{h}}_n \in \mathbb{C}^{N_{\rm t}\times 1}$ can be expressed as follows:
\begin{equation}
	\widetilde{\mathbf{h}}_n=\sum_{p=1}^{N_{\rm p}}\sum_{s=1}^{N_{\rm s}}\alpha_{p,s} \, \mathbf{\rm \mathbf{a}}({\theta_{p,s}}).
\end{equation}
Here, $N_{\rm p}$ represents the number of scattering clusters, $N_{\rm s}$ is the number of sub-paths within each cluster, $\alpha_{p,s}$ denotes the complex gain of each sub-path, and $\theta_{p,s}$ is the angle of departure. The function
$\mathbf{a}(\cdot)$ refers to the steering vector, and $\mathbf{a}({\theta})$ can be mathematically defined as:
\begin{equation}
    \mathbf{a}({\theta})= \left[1,e^{j 2 \pi \frac{\Delta}{\lambda}{\rm sin}(\theta)},\ldots,e^{j 2 \pi \frac{(N_{\rm t}-1)\Delta}{\lambda}{\rm sin}(\theta)} \right],
\end{equation}
where $\Delta$ represents the distance between adjacent antennas, and $\lambda$ is the wavelength. The received signal $y_n \in \mathbb{C}$ on the $n$-th subcarrier is given by:
\begin{equation}
    y_n={\widetilde{\mathbf{h}}_n}^H \mathbf{v}_n x_n + z_n,
\end{equation}
where $\widetilde{\mathbf{h}}_n \in \mathbb{C}^{N_{\rm t}\times 1}$, $\mathbf{v}_n \in \mathbb{C}^{N_{\rm t}\times 1}$, $x_n\in \mathbb{C}$, and $z_n\in \mathbb{C}$ denotes the downlink channel vector, the beamforming vector, the symbol transmitted in the downlink, and the additive noise on the $n$-th subcarrier, respectively. The CSI matrix $\widetilde{\mathbf{H}}$ consists of the stacked channel vectors across $N_{\rm c}$ subcarriers, leading to $\widetilde{\mathbf{H}}=[\widetilde{\mathbf{h}}_1,\ldots,\widetilde{\mathbf{h}}_{N_{\rm c}}] \in \mathbb{C}^{N_{\rm t}\times N_{\rm c}}$. The CSI matrix contains $2N_{\rm t}N_{\rm c}$ real-valued elements, resulting in substantial communication overhead. To enhance sparsity for compression, the CSI matrix is transformed from the time-frequency domain to the angular-delay domain through discrete Fourier transformation (DFT):
\begin{equation}
    \mathbf{H}= \mathbf{F}_{\rm a}\widetilde{\mathbf{H}}\mathbf{F}_{\rm d},
\end{equation}
where $\mathbf{F}_{\rm a} \in \mathbb{C}^{N_{\rm t}\times N_{\rm t}}$ and $\mathbf{F}_{\rm d} \in \mathbb{C}^{N_{\rm c} \times N_{\rm c}}$ are DFT matrices.

\subsection{DL-based CSI feedback}
Considering the CSI matrix $\mathbf{H}$ is overlarge for feedback, a DL-based data compression method is introduced to further alleviate feedback overhead. An autoencoder network is employed, which includes an encoder network deployed at the UE and a decoder network at the BS. The UE first compresses the CSI matrix $\mathbf{H}$ into a low-dimentional codeword ${\rm \mathbf{s}} \in \mathbb{R}^{N_{\rm s}\times 1}$, which can be written as follows:
\begin{equation}
    {\rm \mathbf{s}}=f_{\rm en}(\mathbf{H}),
\end{equation}
where $f_{\rm en}(\cdot)$ represents the encoder network function. The codeword is then fed back to the BS. The compression ratio can be defined as follows:
\begin{equation}
    \gamma=\frac{2N_{\rm t}N_{\rm c}}{N_{\rm s}}.
\end{equation}
The codeword is then transmitted to the BS through a feedback link. The channel of the feedback link is assumed to be perfect. Once receiving the codeword, the BS uses the decoder network to reconstruct the codeword into the CSI matrix, which is formulated as follows:
\begin{equation}
    \mathbf{\widehat{\mathbf{H}}}=f_{\rm de}({\rm \mathbf{s}}),
\end{equation}
where $\mathbf{\widehat{\mathbf{H}}}$ is the reconstructed CSI and $f_{\rm de}(\cdot)$ is the decoder network function. After reconstruction, inverse DFT is performed to convert the CSI matrix back to the spatial-frequency domain.

To implement DL-based CSI feedback, some loss functions are employed to train autoencoder networks. Mean square error (MSE) is the most common loss function in previous related research, which can be formulated as follows:
\begin{equation}
    l_{\rm{MSE}}{\left(\mathbf{H},\widehat{\mathbf{H}}\right)}=\|\widehat{\mathbf{H}}-\mathbf{H}\|_{2}^{2}.
\label{equ:mse}
\end{equation}
Considering that the CSI is usually used for beamforming design, cosine similarity is also taken as a metric of feedback quality \cite{chen2021deep}. In this case, the negative of cosine similarity is employed as a loss function, which can be written as follows:
\begin{equation}
    l_{\rm{CS}}{\left(\widetilde{\mathbf{H}},\widehat{\widetilde{\mathbf{H}}} \right)}=-\frac{1}{N_{\rm c}}\sum_{n=1}^{N_{\rm c}}\frac{|\widehat{\widetilde{\mathbf{h}}}_n^H\widetilde{\mathbf{h}}_n|}{\|\widehat{\widetilde{\mathbf{h}}}_n \|_2 \|\widetilde{\mathbf{h}}_n \|_2}.
\label{equ:cs}
\end{equation}
where $\widehat{\widetilde{\mathbf{h}}}_n$ represents the reconstructed channel of the $n$-th subcarrier.

\section{Communication-efficient Personalized FEEL for DL-based CSI Feedback}

In this section, we introduce a communication-efficient personalized FEEL approach for DL-based CSI feedback. Initially, we present a FEEL-based training framework integrated into DL-based CSI feedback to safeguard data privacy. Recognizing the substantial communication overhead inherent to the FEEL-based framework, we propose a neural network quantization method to alleviate communication burden. Additionally, we introduce a personalization strategy to enhance the performance of DL-based CSI feedback within the FEEL-based training framework.

\subsection{Basic Autoencoder Architecture}

\begin{figure*}[t]
    \centering
    \includegraphics[width=1\textwidth]{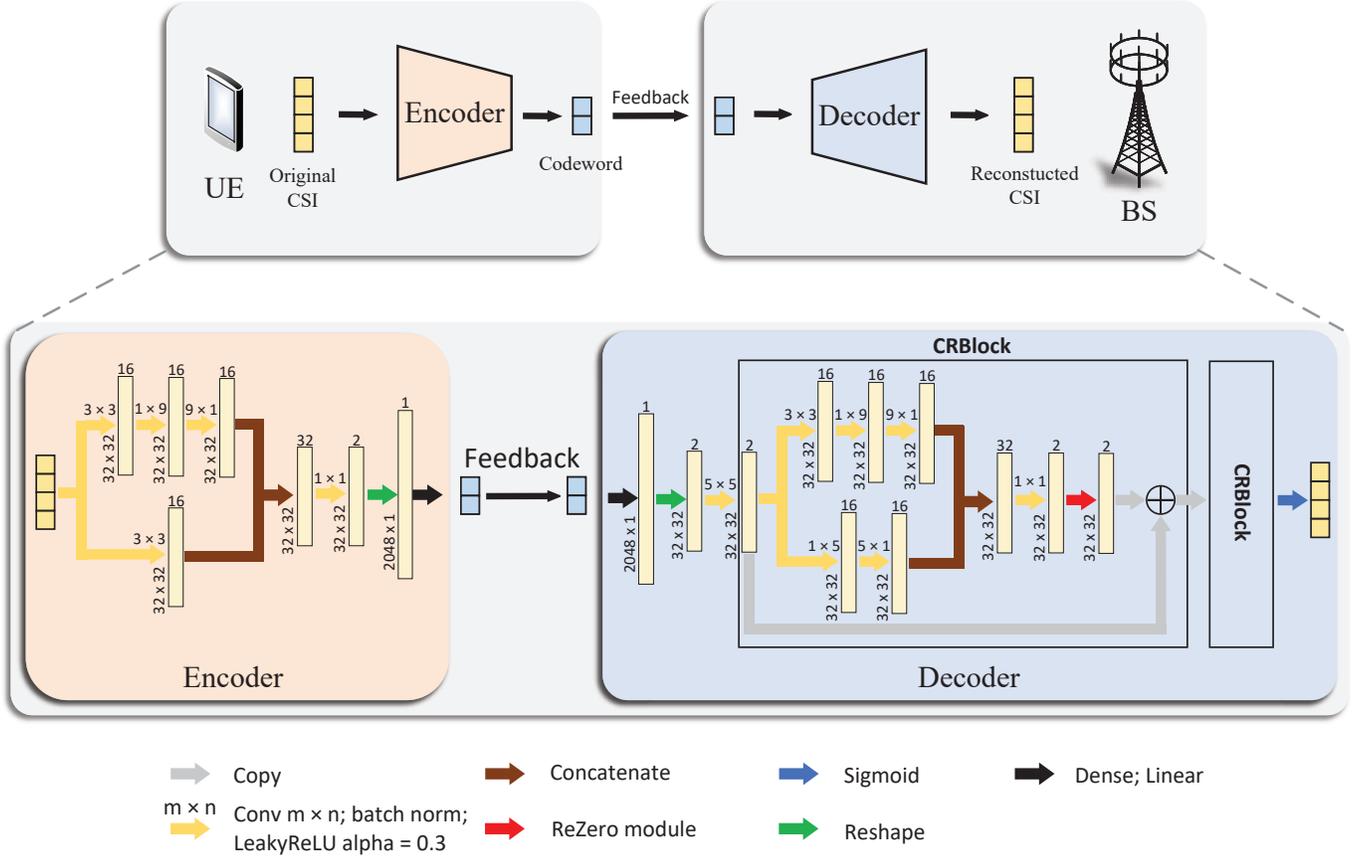}
    \caption{Architecture of the autoencoder network for DL-based CSI feedback.}
    \label{fig:csi}
\end{figure*}

In this study, we employ an adapted version of the channel reconstruction network (CRNet) \cite{lu2020multi} for DL-based CSI feedback.\footnote{Given that CRNet in \cite{lu2020multi} is a lightweight autoencoder and exhibits inferior feedback performance in complex outdoor scenarios, we enhance the performance by widening certain convolutional layers in CRNet.} The architecture of the autoencoder network is illustrated in Figure \ref{fig:csi}. We partition the real and imaginary components of the CSI samples into two separate channels and normalize them to fall within the range of 0 to 1 for input.

Initially, the input CSI undergoes processing through two parallel branches of convolutional layers with distinct kernel sizes, facilitating multi-resolution feature extraction. The resulting features from these branches are concatenated and integrated using a convolutional layer with $1\times 1$ kernels. Following this, the combined features are flattened and compacted into an $N_{\rm s}$-dimensional real-valued vector, referred to as a ``codeword,'' via a dense layer.

Subsequently, at the BS, the codeword is reconstructed by the decoder. This process involves using a dense layer and a reshape layer to restore the original shape of the CSI. A head convolution layer, utilizing $5 \times 5$ kernels, is employed for initial feature extraction. Then, two successive CRBlocks are employed to refine the initially reconstructed CSI, thereby enhancing performance. Each CRBlock adopts a multi-resolution structure akin to the encoder. To circumvent issues related to gradient vanishing and explosion \cite{he2016deep}, we incorporate a residual structure. Furthermore, to expedite convergence, a ReZero module is introduced before the skip connection. This module includes a trainable scalar initialized at zero \cite{bachlechner2021rezero}. Subsequent to the two CRBlocks, a Sigmoid activation is applied to ensure the output values remain within the range of $(0,1)$.

\subsection{FEEL-based Training Framework}
\subsubsection{Motivation}

In conventional CL, the BS can aggregate CSI samples from different UEs across various locations to form a mixed dataset. However, this approach raises concerns of privacy as the location information of UEs can be partially or completely inferred from the collected CSI. Uploading CSI samples to the BS consequently heightens the risk of privacy disclosure. To address these issues, we introduce FEEL to DL-based CSI feedback. FEEL involves local training by each UE and iterative exchange of model parameters with the BS. Crucially, local CSI datasets are not transmitted to the BS, thus preserving UE privacy.

\begin{figure*}[t]
    \centering
    \includegraphics[width=0.80\textwidth]{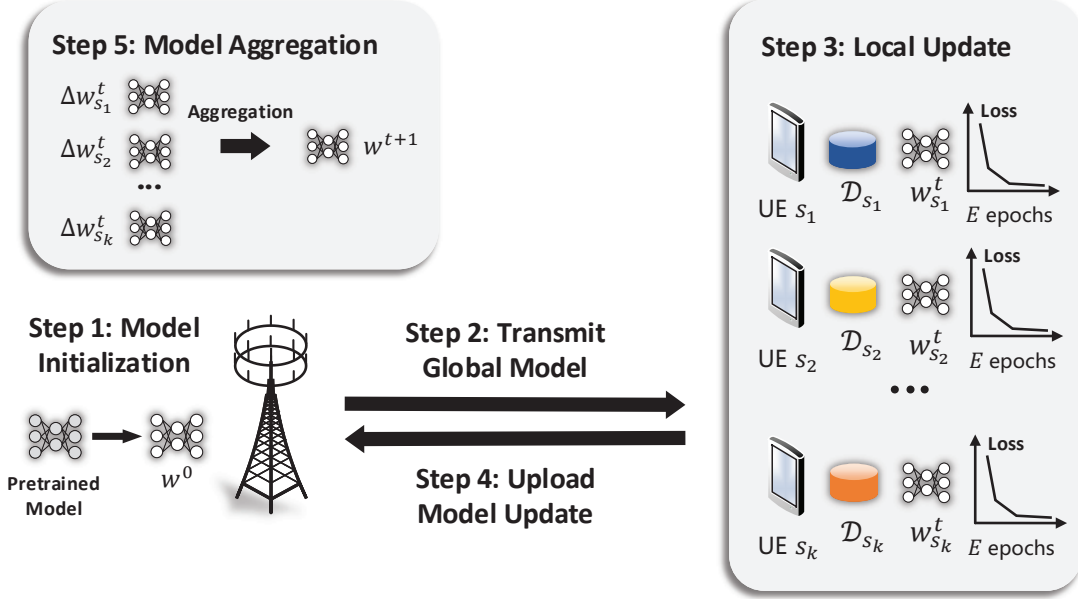}
    \caption{Illustration of the FEEL-based CSI feedback.}
    \label{fig:fl}
\end{figure*}

\subsubsection{Training Framework}

In the FEEL approach, $K$ UEs indexed from $1$ to $K$ collectively participate in training an autoencoder for DL-based CSI feedback. Each UE possesses a local dataset $\mathcal{D}_k$, and the number of samples in $\mathcal{D}_k$ is denoted as $|\mathcal{D}_k|$ with $\sum_{k=1}^{K}|\mathcal{D}_k| = |\mathcal{D}|$. The optimization objective for FEEL can be expressed as:
\begin{equation}
    \min_{w \in \mathbb{R}^{d\times 1}}F(w)\quad {\rm s.t.}\quad F(w)=\sum_{k=1}^{K} \frac{|\mathcal{D}_k|}{|\mathcal{D}|} l_k(w),
    \label{equ:feel}
\end{equation}
where $w$ represents the global model weight, $d$ is the number of network parameters, and $l_k(\cdot)$ denotes the loss on local dataset $\mathcal{D}_k$. The loss function (\ref{equ:mse}) or (\ref{equ:cs}) can be adopted for DL-based CSI feedback tasks. To solve the optimization problem in \eqref{equ:feel}, a FEEL-based training framework \cite{mcmahan2017communication} is proposed for DL-based CSI feedback, as depicted in Fig. \ref{fig:fl}. The key steps of this framework include:

\begin{itemize}
    \item \textbf{Step 1: Model Initialization}. The BS employs a pretrained model $w_0$ to initialize the global model. This pretrained model is often trained using CSI datasets from different scenarios or generated via channel simulation software. Such an approach reduces the training rounds and total communication overhead compared to training from scratch.

    \item \textbf{Step 2: Transmitting Global Model}. The BS assesses UE availability and schedules $M$ available UEs, where $M \leq K$, to participate in the $t$-th round of FEEL. The set of scheduled UEs is denoted as ${\mathcal{S}_t=\{s_1,\ldots,s_{M}\}} \subseteq \{1,\ldots, K\}$. The current global model $w^{t}$ is transmitted to these scheduled UEs for initialization.

    \item \textbf{Step 3: Local update}. Each scheduled UE $s_i$ initializes the local model with the received global model $w^{t}$ and trains it with local CSI datasets $\mathcal{D}_{s_i}$ for $E$ epochs. All UEs adopt consistent training strategies, including learning rate and optimizer choices.

    \item \textbf{Step 4: Uploading Model Update}. After local training, the local model of UE $s_i$ is updated from $w^t$ to $w^t_{s_i}$. Scheduled UEs report the model update $\Delta w^t_{s_i}=w^t_{s_i}-w^t$ to the BS.

    \item \textbf{Step 5: Model Aggregation}. Upon receiving updates $\Delta w^t_{s_i}$ from scheduled UEs, the BS aggregates these updates to generate a new global model $w^{t+1}$. The aggregation is performed by weighting the updates with sample numbers in local datasets $|\mathcal{D}_{s_i}|$. The BS then initiates the subsequent FEEL round from \textbf{Step 2} until the target number of FEEL rounds is reached or the global model meets performance requirements.
\end{itemize}

\begin{algorithm}[t]
	\caption{FedAvg-based FEEL Algorithm for DL-based CSI Feedback}
	\label{alg:FEEL}
	\begin{algorithmic}[1]
		\REQUIRE $K$ UEs indexed as $1,\ldots,K$, each equipped with local CSI datasets denoted as $\mathcal{D}_{1},\ldots,\mathcal{D}_{K}$. In each round of the FEEL process, $M$ UEs are scheduled to participate.  
        The parameters determining the training process are the local epoch number $E$, the batch size $B$, the local learning rate $\eta$, and the loss function $l(\cdot)$.
		\STATE Initialize the global model parameters as $w_0$
		\FOR {$t=1,\ldots,T$}	
		\STATE BS schedules $M$ UEs ${\mathcal{S}_t=\{s_1,\ldots,s_{M}\}} $ and transmits $w^t$ to UEs in $\mathcal{S}_t$.
		\FOR {$ s_i \in \mathcal{S}_t $ in parallel}
        \STATE Initialize $w_{s_i}^t$ as $w_t$
		\STATE Divide $\mathcal{D}_{s_i}$ into batches with batch size $B$
		\FOR {each epoch from $1$ to $E$}
		\FOR {each batch in $\mathcal{D}_{s_i}$}
		\STATE $w_{s_i}^t \leftarrow w_{s_i}^t - \eta \triangledown l_{s_i}(w_{s_i}^t)$ 
		\ENDFOR
		\ENDFOR
		\STATE $\Delta w^t_{s_i} \leftarrow w^t_{s_i}-w^t$
		\STATE Report $\Delta w^t_{s_i}$ to BS
		\ENDFOR
		\STATE $w^{t+1} \leftarrow w^t + \sum_{i=1}^{M} \frac{|\mathcal{D}_{s_i}|}{|\mathcal{D}|} \Delta w^t_{s_i}$
		\ENDFOR
	\end{algorithmic}
\end{algorithm}

The algorithmic representation of the FEEL process is provided in Algorithm \ref{alg:FEEL}. When each UE only performs gradient descent for a single epoch (i.e., $E=1$ in Algorithm \ref{alg:FEEL}), this optimization method is commonly referred to as {\tt FederatedSGD} \cite{mcmahan2017communication}. To expedite the convergence, local updates can be repeated for multiple epochs before transmitting the updates to the BS ($E>1$). This strategy is known as {\tt FederatedAveraging} (FedAvg) \cite{mcmahan2017communication}. In this study, we employ the FedAvg algorithm to enhance communication efficiency.

\subsection{Neural Network Quantization}
\subsubsection{Motivation}
FEEL represents a shift from collecting local datasets to exchanging model parameters (or updates) in comparison to CL. However, the frequent transmission of model parameters presents a significant challenge for FEEL, particularly in the context of DL-based CSI feedback. Autoencoder networks typically possess a large number of parameters, leading to excessive communication overhead.

Additionally, transmitting model updates in the uplink direction poses an even greater challenge. Firstly, there exists an asymmetry in bandwidth between uplink and downlink transmission, with uplink transmission speeds usually being significantly slower than downlink. Moreover, as highlighted by \cite{konevcny2016federated}, applying certain cryptographic protocols to enhance privacy during UE model update uploads results in additional communication overhead. To tackle these issues and enhance communication efficiency, the FEEL-based training framework incorporates neural network quantization. 

\subsubsection{Key Idea}

Autoencoder network parameters are typically stored as 32-bit floating-point numbers, which is redundant for representing the network. To reduce communication overhead, this work employs parameter quantization for autoencoder networks, which involves representing parameters using fewer bits \cite{choudhary2020comprehensive}. 

In an autoencoder network, the quantity and impact on performance vary for different types of parameters. For example, the parameters of a dense layer or a convolutional layer comprise weights and biases. Weight parameters typically constitute a significant portion of the total parameters and are less sensitive to quantization errors. Conversely, bias parameters are fewer in number and more susceptible to quantization errors. Consequently, only weight parameters are quantized, while bias weights are still stored as 32-bit floating-point numbers. The quantization of batch normalization and ReZero layers has minimal impact on the total parameter count and is not considered.

The effect of quantization on uplink and downlink transmission is then analyzed. During uplink transmission, UEs transmit model updates to the BS. The updates from all UEs are aggregated by averaging at the BS, which can mitigate the impact of quantization noise to some extent. As a result, the quantization errors in uplink transmission have a relatively small influence on feedback performance, allowing for lower precision quantization of the model update. In contrast, for downlink transmission, the BS transmits the global model to the UEs. The global model is directly used to initialize the autoencoder network for local updates, making it more susceptible to quantization errors. Consequently, more bits are required to represent the global model to minimize the impact of quantization compared to the uplink model updates.  

\subsection{A Fine-tuning-based Personalization Strategy}

\begin{figure}[t]
    \centering
    \includegraphics[width=0.5\textwidth]{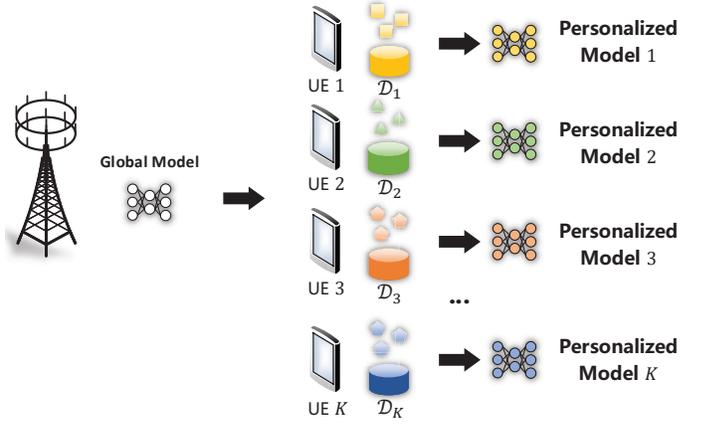}
    \caption{Illustration of the fine-tuning-based personalization strategy for FEEL-based CSI feedback.}
    \label{fig:pfl3}
\end{figure}

\subsubsection{Motivation}
Data heterogeneity presents a significant challenge in FEEL, as it contradicts the common assumption of all training data being independently and identically distributed (i.i.d.) in typical machine learning problems \cite{mcmahan2017communication}. In DL-based CSI feedback, the channel characteristics of UE-BS links predominantly depend on the physical environments surrounding UEs. For UEs located in different positions within a cell, their channel characteristics can vary substantially. Such data heterogeneity can lead to performance degradation and poor convergence in FEEL. Notably, previous studies on FEEL have not always showcased superior performance over individual learning (IL), especially when each UE possesses sufficient CSI samples for training. This unfavorable performance might discourage UEs from participating in FEEL.

As indicated in \eqref{equ:feel}, the objective of FEEL is to simultaneously optimize the feedback performance of a large number of UEs in a cell. After optimization, the autoencoder converges to an average solution that maximizes performance across all UEs. This solution is well-generalized to all UEs but may still be suboptimal from the perspective of each individual UE. On the other hand, IL aims to optimize the feedback performance of each UE in isolation, disregarding the performance of the autoencoder on other UEs. Consequently, IL often achieves relatively high performance but lacks favorable generalization. An intuitive approach is to combine the strengths of FEEL and IL, leveraging the benefits of both methods. To integrate the advantages of IL into FEEL, the global model of FEEL can serve as a well-generalized initialization and then undergo retraining using local CSI datasets, similar to IL. Additionally, as highlighted by \cite{jiang2019improving}, FEEL shares substantial similarities with a meta-learning method named Reptile \cite{nichol2018first}, which can naturally provide an initialization suitable for fine-tuning. Based on this concept, a fine-tuning-based personalization strategy is proposed for FEEL 

\subsubsection{Training Framework}
In traditional FEEL, all UEs share a global model for DL-based CSI feedback tasks, and the optimization objective can be expressed as in \eqref{equ:feel}. In contrast, personalized FEEL enables each UE to optimize a personalized model to adapt to its local CSI dataset. The objective can be reformulated as follows: 
\begin{equation}
\begin{aligned}
    \min_{w_1,\ldots, w_K \in \mathbb{R}^{d\times 1}}&F(w_1,\ldots,w_K) \quad \\
    {\rm with}~~~~&F(w_1,\ldots,w_K) = \sum_{k=1}^{K} \frac{|\mathcal{D}_{k}|}{|\mathcal{D}|} f_k(w_k).
    \label{equ:pfeel}
\end{aligned}
\end{equation}

\begin{figure*}[t]
    \centering
    \includegraphics[width=0.80\textwidth]{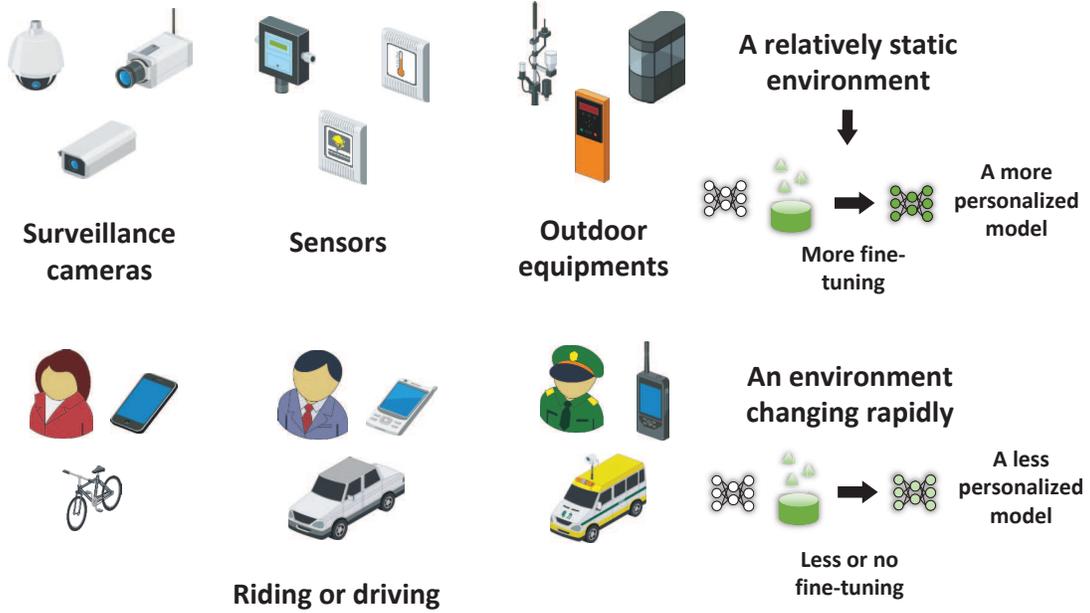}
    \caption{Application of personalization in practical scenarios.}
    \label{fig:pfleg}
\end{figure*}

The personalization strategy is depicted in Fig. \ref{fig:pfl3}. Once the global model attains the desired performance, the BS concludes the next iteration of FEEL and transmits the current global model to all UEs. Each UE can then commence fine-tuning the global model using its local dataset, resulting in each UE possessing a distinct personalized model that is better suited for CSI compression and reconstruction in its specific environment. While searching for optimal freezing strategies during the fine-tuning procedure may be useful, it can vary based on autoencoder architectures and the number of samples, leading to deployment complexity. Therefore, in this study, all parameters are fine-tuned, representing a straightforward yet effective approach for most cases. Subsequently, after personalization, since the strategy is executed on the UE side, all UEs need to transmit the personalized decoder networks back to the BS to replace the global decoder. Furthermore, after personalization, the feedback performance of personalized models is monitored by UEs using the latest CSI samples, as UEs may move to different environments. If the performance of a personalized model deteriorates compared to the global model, the personalized model is discarded, and the global model is reinstated.

An important trade-off exists between generalization capability and feedback performance in personalized FEEL. During fine-tuning, the feedback performance of a specific UE increases, while its generalization capability diminishes. As illustrated in Fig. \ref{fig:pfleg}, UEs can flexibly choose the number of fine-tuning epochs to determine whether to prioritize a more generalized model or a high-performance model, based on their environmental conditions. For instance, in scenarios involving highly mobile UEs, such as driving scenarios, rapid environmental changes make achieving better generalization challenging. Consequently, no fine-tuning or only a limited number of fine-tuning epochs can be employed to enhance generalization. Conversely, for IoT devices (e.g., surveillance cameras and sensors) or UEs confined to a small area, more fine-tuning epochs can be adopted to further enhance feedback performance.

\section{Simulation Results}
In this section, we present the channel generation settings and the training strategies. Subsequently, we analyze the performance of FEEL for DL-based CSI feedback and the quantization method. Lastly, we evaluate the proposed personalization strategy. 

\subsection{Simulation Settings}

\subsubsection{Channel Generation}
The CSI datasets are generated using QuaDRiGa \cite{jaeckel2014quadriga}, a geometry-based stochastic channel model generator.\footnote{\url{https://quadriga-channel-model.de}} We consider two scenarios: RMa NLOS (Non-Line-of-Sight) for pretraining and UMi LOS (Line-of-Sight) for deployment. These scenarios correspond to the {\tt 3GPP\_38.901\_RMa\_NLOS} and {\tt 3GPP\_38.901\_UMi\_LOS} presets in QuaDRiGa, compliant with 3GPP TR 38.901 \cite{38901}. The choice of scenarios enables us to simulate the case where the distribution of CSI in a practical scenario differs from that in the dataset collected or generated for pretraining. The cell range is set to 100 m for both scenarios. The center frequency and bandwidth are 2.655 GHz and 70 MHz, respectively, corresponding to the operating band ``n7'' as defined in 3GPP TS 38.101-1 \cite{38101}. Specific settings are listed in Table \ref{tab:my_label}.

\begin{table}[t]
    \renewcommand{\arraystretch}{1.0}
    \centering
    \caption{Channel generation settings}
    \begin{tabular}{|l|l|} 
    \hline
        \multirow{2}{*}{\textbf{Scenarios}} & Pretraining: RMa NLOS \\
         & Deployment: UMi LOS\ \\
         \hline
         \multirow{2}{*}{\textbf{Antenna settings}}&BS: 32 omnidirectional antennas, ULA \\
         & UE: a single omnidirectional antenna \\
         \hline
         \textbf{Operating system} & FDD-OFDM system \\
         \hline
         \textbf{Subcarrier number} & 32 \\
         \hline
         \textbf{Center frequency} & 2.655 GHz \\
         \hline
         \textbf{Bandwidth} & 70 MHz \\
         \hline
         \textbf{Cell range} & 100 m \\
         \hline
         \textbf{BS height} & RMa: 25 m \quad UMi: 10 m \\
         \hline
         \textbf{UE height} &  1.5 m \\
         \hline
         \textbf{Minimum BS-UE distance} & 10 m \\
         \hline
         \textbf{Correlation distance} & RMa: 50 m \quad UMi: 12 m \\
    \hline 
    \end{tabular}
    \label{tab:my_label}
\end{table}

\begin{figure}[t]
    \centering
    \includegraphics[width=0.45\textwidth]{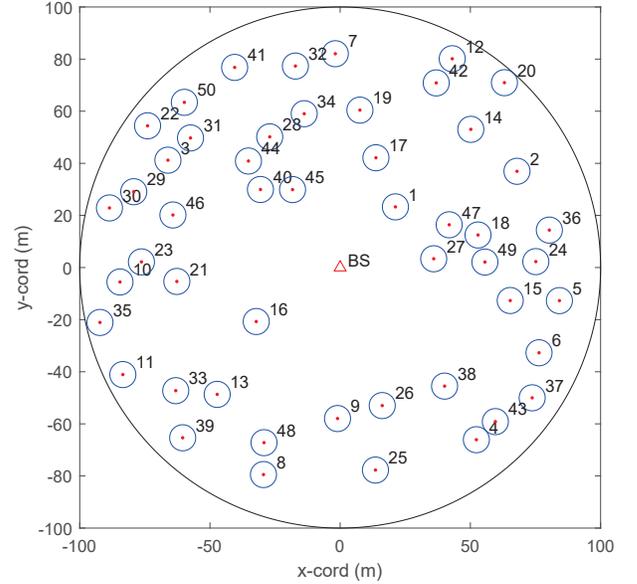}
    \caption{UE positions in the RMa and UMi scenarios. The black circle represents the cell boundary, and the blue circles indicate the range of movement for the UEs.}
    \label{fig:umi}
\end{figure}

Figure \ref{fig:umi} illustrates the generation of 50 UEs, indexed from 1 to 50, in the RMa and UMi scenarios. The BS is located at the center of the cell, and the UEs' positions are uniformly distributed throughout the cell. Each UE is allowed to move randomly within a circular area with a radius of 5 m, and 10,000 CSI samples are generated in this area for each UE. The CSI datasets of all UEs are divided into training, validation, and test sets with an 8:1:1 split.\footnote{In practical deployment, the BS requires validation and test datasets to monitor the FEEL process and select the best model trained by FEEL. The BS is assumed to possess a small amount of CSI datasets from different UEs used for validation and testing.} All CSI samples are normalized to the range of 0 to 1.

\subsubsection{NN Training Details}
All DL-based algorithms were implemented using TensorFlow 2.4.1 on a DGX-1 workstation. The learning rate was initialized at $10^{-3}$ and reduced to $10^{-4}$ when the loss function converged. A batch size of 32 was used, and the Adam optimizer was employed for both FEEL and personalization. In each FEEL round, 5 UEs were randomly selected from all UEs to participate. The number of local epochs was set to 2. The normalized MSE (NMSE) was used to evaluate the performance of CSI feedback: 
\begin{equation}
	\rm NMSE= \rm E\left\{ {\|\mathbf{H}-\widehat{\mathbf{H}}\|_2^2}/{\|\mathbf{H}\|_2^2}\right\}.
\end{equation}

\begin{figure}[t]
    \centering
    \includegraphics[width=0.45\textwidth]{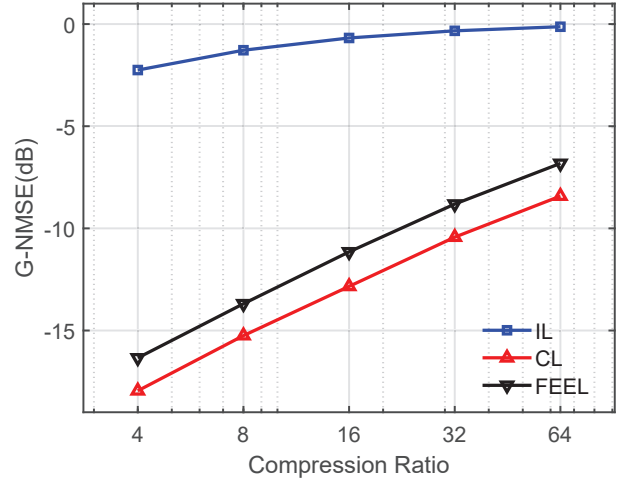}
    \caption{Comparison of IL, CL, and FEEL for DL-based CSI feedback under different compression ratios. The number of UEs and samples per UE are 50 and 5,000, respectively.}
    \label{fig:feelcr}
\end{figure}

Two metrics were used to evaluate the FEEL-based training framework and the personalization strategy based on NMSE. For the FEEL-based training framework, the global NMSE (G-NMSE) was calculated on the mixed datasets of all 50 UEs. This metric represents the feedback performance when a UE moves randomly within the cell and can also be interpreted as the generalization capability of a model. On the other hand, for the personalization strategy, the performance of the personalized models on the corresponding UE was evaluated. Therefore, the individual NMSE (I-NMSE) was also adopted as a metric, using the average NMSE on each individual UE.

Four different training frameworks were simulated for DL-based CSI feedback for comparison: IL, CL, FEEL, and personalized FEEL. IL refers to the independent training of an autoencoder network by each UE using their respective CSI datasets. CL refers to the training of an autoencoder by the BS with the mixed datasets collected from all UEs. FEEL and personalized FEEL follow the descriptions provided in Section III.

\subsection{FEEL Performance for DL-based CSI Feedback}

\begin{figure}[t]
    \centering
    \includegraphics[width=0.45\textwidth]{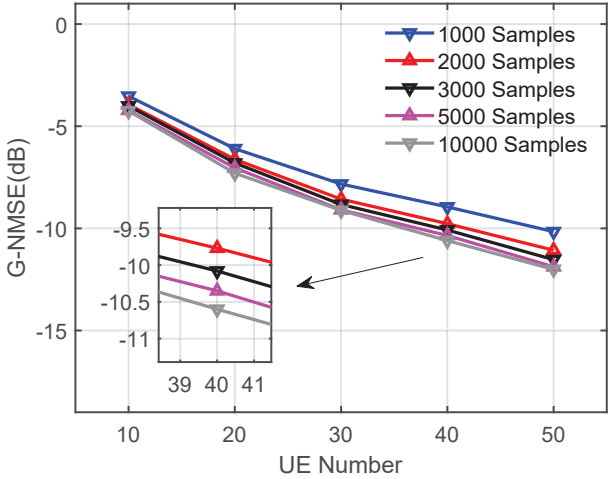} 
    \caption{Performance of FEEL for DL-based CSI feedback under different numbers of UEs and samples per UE. The compression ratio is set to 1/16.}
    \label{fig:feelue}
\end{figure}

The performance comparison of different training frameworks, namely IL, CL, and FEEL, is presented under various compression ratios. As shown in Fig. \ref{fig:feelcr}, the FEEL-based training framework demonstrates comparable performance to CL, showcasing its effectiveness for DL-based CSI feedback. The small performance gap between FEEL and CL can be attributed to the non-i.i.d. nature of local CSI datasets, a gap that advanced scheduling strategies can help mitigate \cite{xia2020multi,yang2020scheduling}. Moreover, both FEEL and CL outperform the IL-based training framework significantly. The autoencoder networks trained through CL and FEEL exhibit favorable generalization capabilities across different environments, whereas those trained through IL fail to generalize to new channel conditions. This highlights the importance of UE collaboration in DL-based CSI feedback.

\begin{figure}[t]
    \centering
    \includegraphics[width=0.45\textwidth]{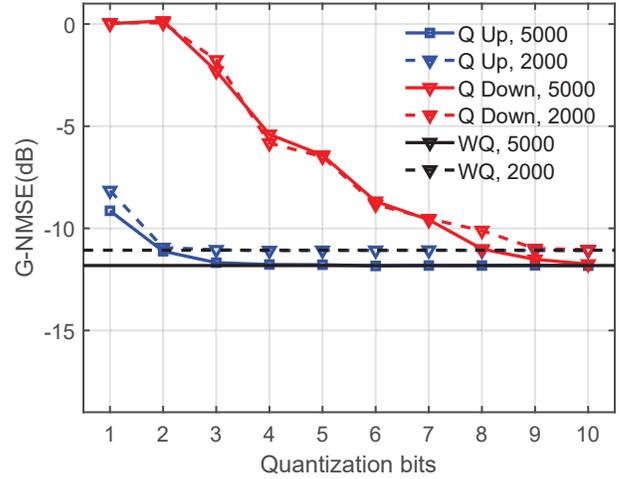}
    \caption{Performance of FEEL for DL-based CSI feedback under different quantization bits. The compression ratio and UE number are set to 1/16 and 50, respectively. `Q', `WQ', `Up', `Down' represent quantized FEEL, FEEL without quantization, uplink quantization, and downlink quantization, respectively. The number in the legend denotes the number of samples per UE.}
    \label{fig:quan}
\end{figure}

In real-world deployment, UEs often have limited storage capacity and may not accumulate enough CSI samples for FEEL. Also, unstable connections or other bandwidth-intensive applications might limit UEs' participation in FEEL. Therefore, the impact of sample and UE numbers is investigated. Fig. \ref{fig:feelue} illustrates feedback performance for varying sample numbers per UE and UE numbers. The autoencoder network's feedback performance improves as more UEs engage in FEEL, indicating that increased participation enhances feedback performance by incorporating more diverse channel observations in training. However, the sample numbers have a relatively minor influence on FEEL. When each UE has 3,000 CSI samples, the G-NMSE only drops by less than 1 dB compared to the scenario with 10,000 samples per UE. This suggests that even with a small number of training samples per UE participating in FEEL, the collaborative nature of the process plays a role akin to data augmentation. Hence, the collective training samples of all UEs are usually adequate, and individual UEs do not require a large storage capacity.

\begin{figure}[t]
    \centering
    \includegraphics[width=0.45\textwidth]{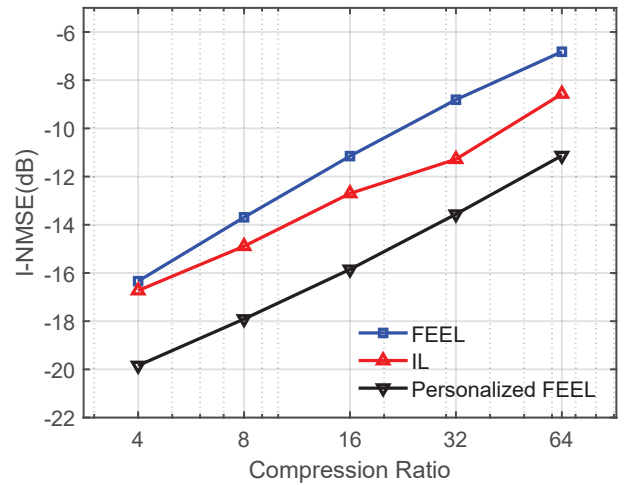}
    \caption{Comparison of IL, FEEL, and personalized FEEL for DL-based CSI feedback under different compression ratios. The number of UEs and samples per UE are 50 and 5,000, respectively.}
    \label{fig:pfl}
\end{figure}

The performance of the FEEL-based training framework using neural network quantization is then evaluated. Different quantization bit settings are tested to determine the optimal number of bits for both uplink and downlink transmission. This is done to reduce communication overhead while maintaining a relatively low performance drop. Fig. \ref{fig:quan} depicts the evaluation results. It is evident that a minimum of 2 quantization bits is needed when quantizing only the uplink model updates to maintain performance, resulting in less than 1 dB performance drop compared to non-quantized FEEL. When considering quantization of downlink global models, at least 8 bits are required to achieve comparable performance with non-quantized FEEL. The disparity between the two quantization strategies mainly stems from the aggregation process at the BS, which partially alleviates the impact of quantization noise in the uplink. With a 2-bit quantization for uplink and 8-bit quantization for downlink, communication overhead in the FEEL-based training framework can be reduced by approximately 16 times for uplink and 4 times for downlink, substantially enhancing communication efficiency. The impact of quantization bits is also assessed with varying sample numbers per UE, revealing that sample numbers have minimal influence on the proposed quantization strategy.

\subsection{Performance of Personalized FEEL for DL-based CSI Feedback}

In Fig. \ref{fig:pfl}, the I-NMSE of IL, FEEL, and personalized FEEL are compared under different compression ratios to assess the individual performance of autoencoder networks. It is evident that personalized FEEL outperforms FEEL, demonstrating that the proposed personalization strategy effectively adapts the global model trained by FEEL to the specific environment of each UE using local datasets. Notably, the simulation considers a worst-case scenario where UEs randomly move within certain ranges. In real-world scenarios, UEs often follow regular paths, such as walking along streets, where the performance gain due to the personalization strategy could be even greater. Additionally, personalized FEEL consistently outperforms IL across various compression ratios, as FEEL can leverage shared environmental knowledge from other UEs to provide a better initialization for personalization compared to IL.

\begin{figure*}[t]
	\centering
	\subfigure [\label{fig_first_case1} ]{
		\includegraphics[width=0.45\linewidth]{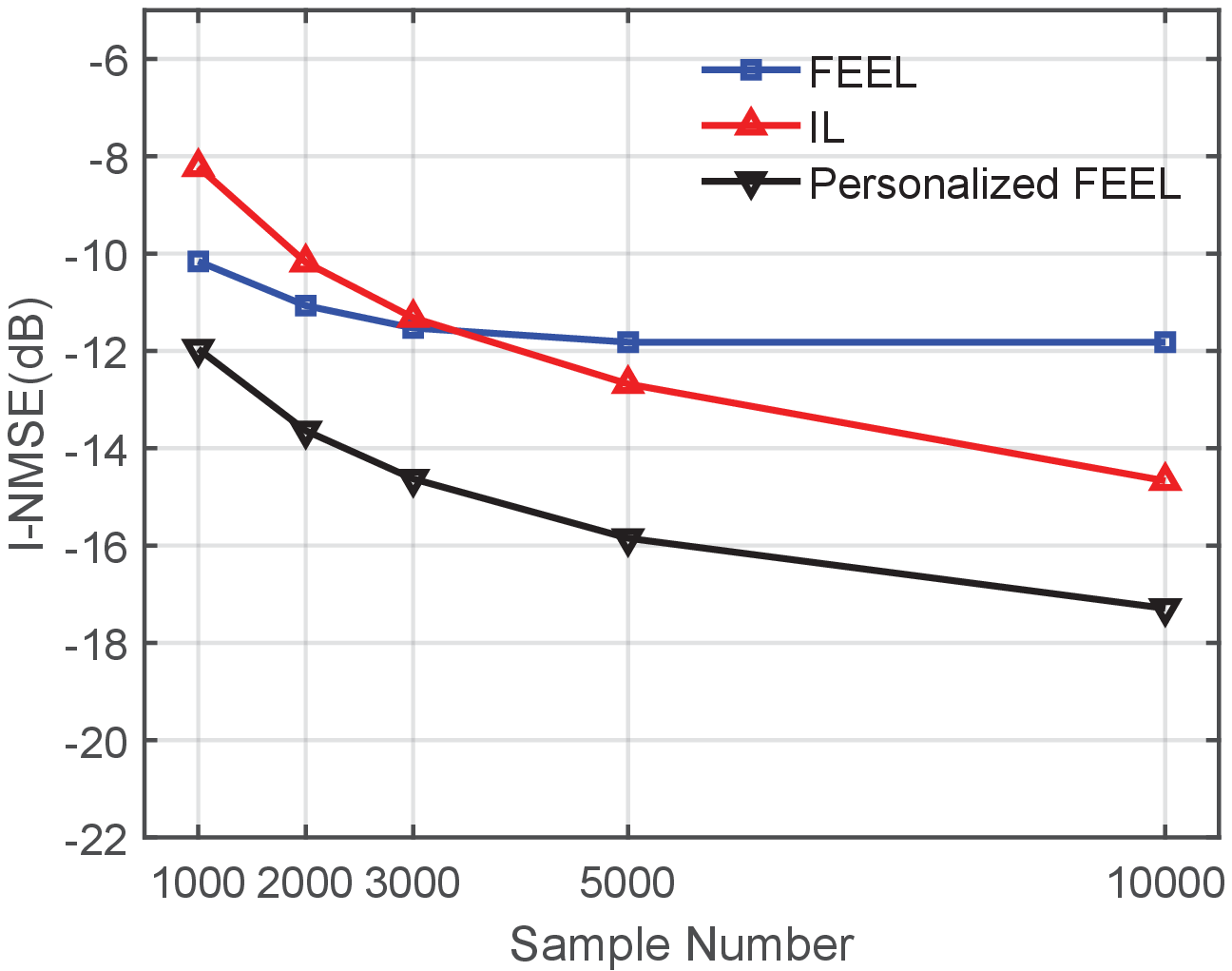}}
	\subfigure [\label{fig_second_case1}]{
		\includegraphics[width=0.45\linewidth]{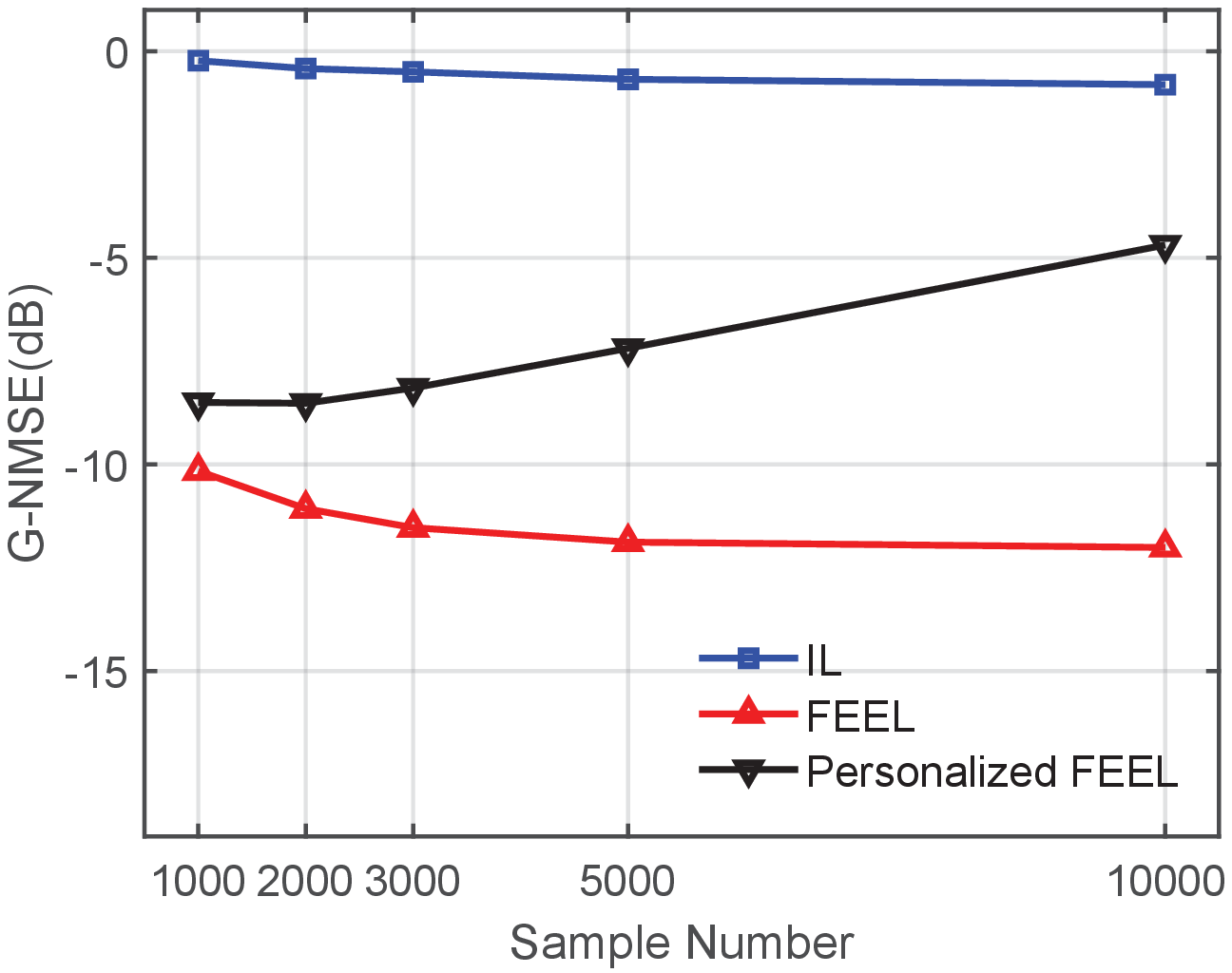}}	
		\caption{Comparison of feedback performance and generalization capability of IL, FEEL, and personalized FEEL under different sample numbers. The compression ratio and UE number are 1/16 and 50, respectively.}
	\label{fig:feelsample}
\vspace{-0.7cm}
\end{figure*}

In a similar vein, Fig. \ref{fig:feelsample}(a) explores the influence of sample numbers on personalized FEEL, taking into account the limited storage capacity of UEs. With 50 UEs participating in FEEL, performance remains relatively stable with varying sample numbers due to the collaboration among a large number of UEs. However, the performance of IL is sensitive to sample numbers and significantly drops with insufficient training samples. In these settings, IL requires approximately 3,000 training samples to surpass FEEL. This highlights how FEEL can enhance feedback performance through collaboration even when training samples are scarce. Moreover, personalized FEEL consistently shows performance gains across different sample numbers when compared to FEEL and IL. With each UE having 10,000 training samples, personalized FEEL's performance gain exceeds 5 dB compared to FEEL. Increasing the number of training samples for fine-tuning further boosts these performance gains.

Considering that a UE launching personalization in one environment might move to another environment, the generalization capability of personalized FEEL is tested across different sample numbers, as depicted in Fig. \ref{fig:feelsample}(b). The generalization capability of personalized models decreases as more training samples are used for fine-tuning. However, the generalization capability of the personalized models remains much higher than that of models trained by IL.

\begin{figure*}[t]
	\centering
	\subfigure [\label{fig_first_case2} ]{
		\includegraphics[width=0.45\linewidth]{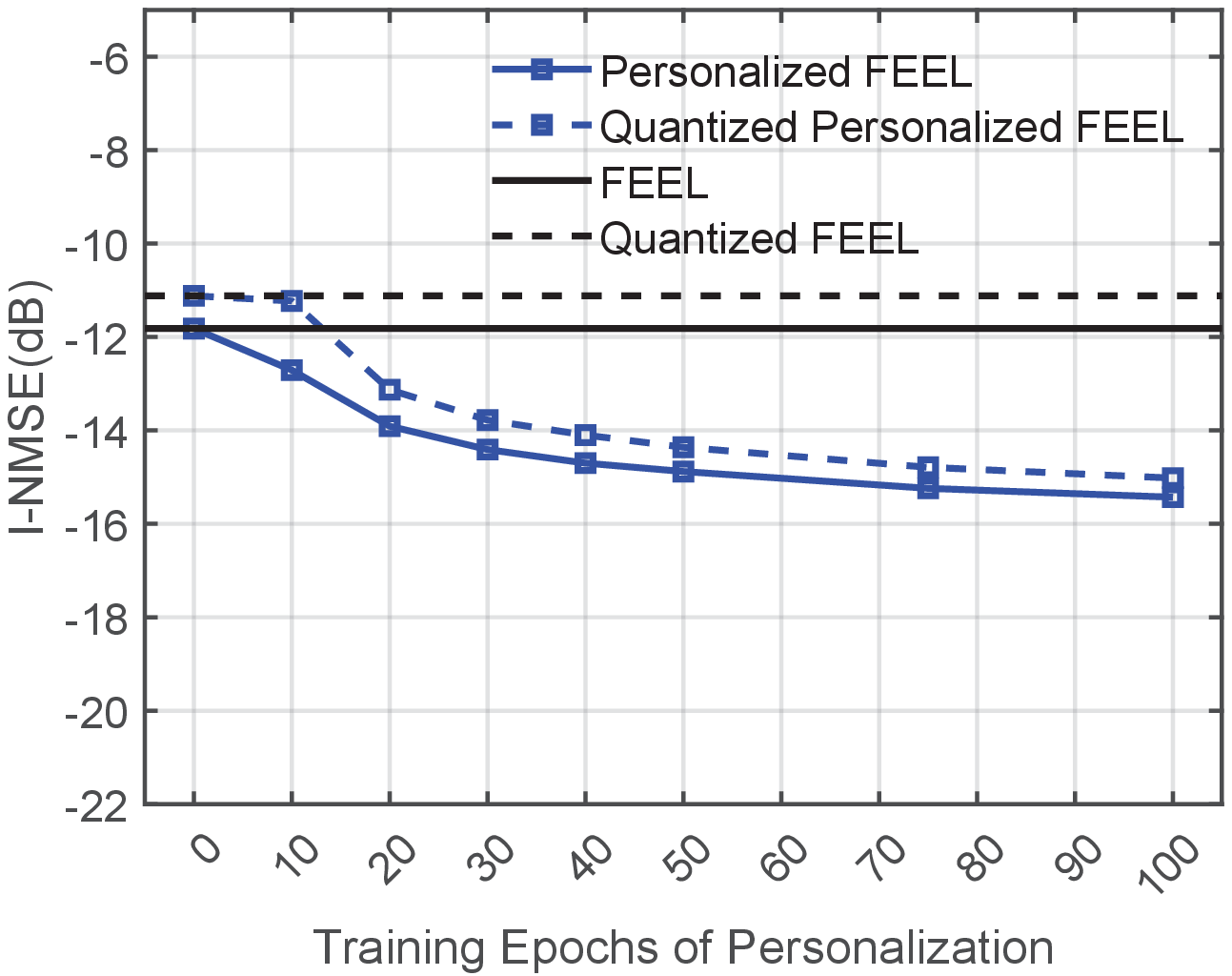}}
	\subfigure [\label{fig_second_case2}]{
		\includegraphics[width=0.45\linewidth]{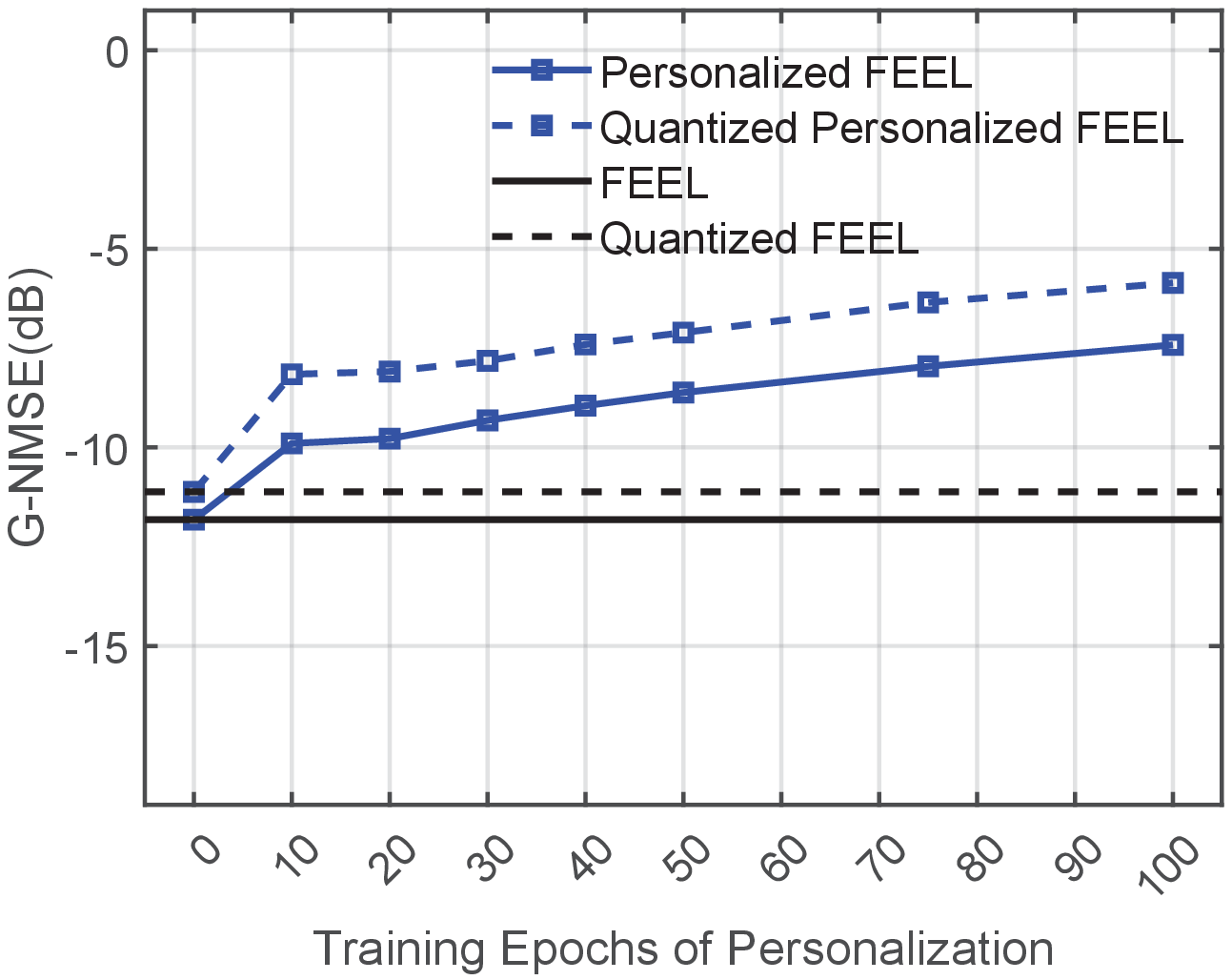}}   	
	\caption{Performance-generalization tradeoff in personalized FEEL without quantization. The compression ratio, number of UEs, samples per UE, and quantization bits are set to 1/16, 50, 5,000, and 2, respectively.}
	\label{fig:tradeoff}
\vspace{-0.7cm}
\end{figure*}

The tradeoff between performance and generalization in personalized FEEL is shown in Fig. \ref{fig:tradeoff}.\footnote{The stepwise learning rate strategy mentioned earlier is challenging to implement when the training epoch of personalization is fixed. For this simulation, the learning rate is set at a fixed value of $10^{-3}$.} During personalization, the I-NMSE decreases as the training epochs increase, indicating improved feedback performance for each UE. However, the G-NMSE increases during personalization, signifying a decline in the generalization capability of the personalized models with more fine-tuning epochs. Despite this reduction, the G-NMSE remains lower than -5 dB, which is significantly superior to the IL-based training framework (G-NMSE = -0.68 dB). In practical scenarios, UEs can select an appropriate number of epochs based on their surrounding environments. Moreover, the impact of quantization on personalization is evaluated. In this evaluation, a global model with 2-bit quantization in the uplink is employed for personalization. The results indicate that feedback performance only slightly degrades when personalization is conducted on a quantized global model.

\begin{figure}[t]
    \centering
    \includegraphics[width=0.45\textwidth]{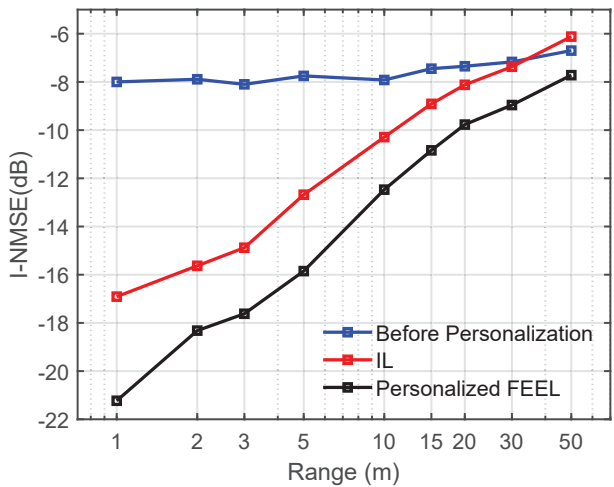}
    \caption{Performance of personalized FEEL for DL-based CSI feedback with different ranges of UEs. The compression ratio, number of UEs, and samples per UE are set to 1/16, 50, and 5,000, respectively.}
    \label{fig:pflrange}
\end{figure}

In practical deployment, the trajectories of UEs can vary widely based on their characteristics and activities, leading to differing performance gains from the personalization strategy. To explore the impact of UE movement on personalized FEEL, an investigation into the influence of the moving range of UEs is conducted. For each moving range, 50 distinct UEs are generated within the same cell, with each UE possessing 5,000 CSI samples. As depicted in Fig. \ref{fig:pflrange}, the results consistently indicate that personalized FEEL consistently outperforms IL, regardless of the range within which UEs move. Furthermore, it is observed that the feedback performance of personalized models tends to increase with smaller moving ranges. Specifically, when UEs remain within relatively static environments (e.g., a moving range of 1 m), the personalization gain based on the global model can reach as high as 13.23 dB.\footnote{ It is important to note that the environments of the UEs participating in FEEL are different from those of the UEs used for personalization in this simulation. This leads to the degradation of the I-NMSE of the global model compared to the previous simulations. However, simulations indicate that this degradation can be effectively mitigated (less than 1 dB) when the number of participating UEs in FEEL exceeds 200.} Importantly, scenarios where UEs are confined to smaller areas (e.g., with moving ranges smaller than 5 m) can also be seen as non-worst cases compared to the results presented in Fig. \ref{fig:pfl}, further underscoring the high potential of the personalization strategy. Moreover, the results suggest that the personalization strategy is particularly effective for UEs that remain within limited areas. Interestingly, even in scenarios where UEs move over relatively larger ranges (e.g., a moving range of 30 m), the personalization strategy still leads to significant performance gains. This implies that the personalization strategy can be applied effectively in various real-world conditions, as most situations can be considered subsets of the scenario where UEs move randomly within a circular area with a radius of 30 m.
 
\subsection{FEEL and Meta-learning}

\begin{figure}[t]
    \centering
    \includegraphics[width=0.45\textwidth]{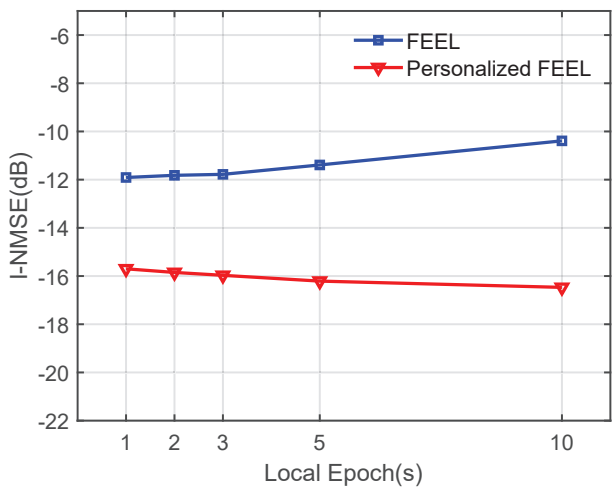}
    \caption{Performance of FEEL and personalized FEEL for DL-based CSI feedback with different numbers of local training epochs. The compression ratio, number of UEs, and samples per UE are set to 1/16, 50, and 5,000, respectively.}
    \label{fig:feele}
\end{figure}

In this subsection, we delve into the influence of the number of local training epochs within the FEEL framework. The performance of both FEEL and personalized FEEL are evaluated while varying the number of local epochs. The results depicted in Figure \ref{fig:feele} reveal interesting trends: the performance of personalized FEEL improves as the number of local epochs increases, while FEEL's performance declines as each UE conducts more local epochs. This suggests that the benefits of personalization become more pronounced when longer local training is performed within each iteration of FEEL.

This phenomenon can be attributed to the underlying similarity between the FEEL training process and a first-order meta-learning algorithm known as Reptile \cite{nichol2018first}. As evidenced by previous research \cite{jiang2019improving,nichol2018first}, the FEEL training process amalgamates joint learning and meta-learning aspects. When each UE trains its local model for fewer epochs, FEEL aligns more closely with joint learning, optimizing performance across all participating UEs and leading to improved overall FEEL performance. Conversely, when local training is extended to more epochs, FEEL resembles meta-learning more closely, optimizing the convergence of subsequent fine-tuning in personalization and ultimately yielding higher performance for personalized FEEL.

\section{Conclusion}

In this paper, we introduce a novel communication-efficient personalized FEEL-based training framework, presenting an innovative approach to enhance the classic DL-based CSI feedback algorithm. The proposed method offers a substantial advantage by enabling the autoencoder network to achieve high performance while maintaining moderate generalization. Firstly, we propose a FEEL-based training framework that not only preserves UE privacy but also enhances feedback performance. Secondly, a neural network quantization method is introduced to minimize the communication overhead in FEEL. Lastly, we present a personalization strategy aimed at further improving feedback performance. The effectiveness of these methods is evaluated using channel datasets generated by QuaDRiGa.

The results illustrate that our proposed methods can enhance feedback performance while retaining a reasonable level of generalization capability, all while exhibiting high communication efficiency. Moreover, these benefits are particularly pronounced in scenarios where UEs remain in relatively static environments. Additional improvements are observed with larger sample sizes per UE and increased local training epochs.

We anticipate that our work will stimulate future research in the realm of personalized FEEL-based CSI feedback, particularly in the following directions: 
\begin{itemize}
    \item \textbf{Unreliable connections}: In real-world scenarios, the connection between UEs and BS might not always be reliable enough to support the FEEL-based training framework. Exploring methods to mitigate the degradation of feedback performance caused by transmission errors is a promising avenue of research.
     
    \item \textbf{Quantification of privacy gain}: While FEEL is acknowledged for its privacy-preserving attributes compared to CL, quantifying the privacy gain of the FEEL-based training framework, especially in the context of DL-based CSI feedback, remains a challenge. Developing methods to quantitatively evaluate the privacy advantages of FEEL would enhance the credibility of its benefits. 
        
    \item \textbf{Hybrid-field channels}: The future of cellular networks might involve extremely large antenna arrays and the use of higher frequency bands, potentially leading to the coexistence of near-field and far-field conditions within the same cell. While DL-based CSI feedback can adapt to varying channel distributions, the heterogeneity of fields might also impact the FEEL-based training framework's performance. Addressing this issue is a significant challenge worth exploring. 
\end{itemize}

\bibliographystyle{IEEEtran}
\bibliography{refer1}

\end{document}